\documentstyle[psfig]{elsart}

\newcommand{\delone}{\mbox{$\Delta I\!\!=\!\!1$ }}
\newcommand{\deltwo}{\mbox{$\Delta I\!\!=\!\!2$ }}
\begin{document}
\begin{frontmatter}
\title{\large \bf Wobbling motion in the multi-bands crossing region}
\author[tokyo,wako]{Makito Oi} 
\footnote{e-mail address: mon@cns.s.u-tokyo.ac.jp},
\author[tokyo,india]{Ahmad Ansari},
\author[aomori,wako]{Takatoshi Horibata}
\author[tokyo,wako]{Naoki Onishi}
\address[tokyo]{Institute of Physics, Graduate School of Arts and Sciences,
University of Tokyo, Komaba, Meguro-ku, Tokyo 153-8902, Japan}
\address[india]{Institute of Physics, Doordarshan Marg, 
Bhubaneswar 751 005, India}
\address[aomori]{Department of Information System Engineering,
Aomori University, Kobata, Aomori 030-0943, Japan}
\address[wako]{Cyclotron Laboratory, Institute of Physical and Chemical
Research (RIKEN),Hirosawa 2-1, Wako-city, Saitama 351-0198, Japan}

\begin{abstract}
The backbending in the $A\simeq 180$ mass region is expected to be 
caused by multi-bands crossing between low-$K$ (g- and s-bands) 
and high-$K$ bands ($K^{\pi}\!\!=\!\!8^+$ or $\!\!10^+$).
We analyze a mechanism of coupling of these bands
in terms of a dynamical treatment for nuclear rotations, i.e.,
the wobbling motion. The wobbling states are produced through the
generator coordinate method after angular momentum projection,
in which the intrinsic states are constructed through the 2d-cranked
HFB calculations.
\end{abstract}
\end{frontmatter}

A new interest in the ``backbending'' phenomenon has arisen 
through a series of experiments in 1990s' \cite{WY93,KL95,SM95,SM98}  
for nuclei in the $A\simeq 180$ region.
It is expected that 
the backbending in this mass region is caused 
by the three-bands crossing, i.e., the ground (g-), super (s-) 
and ``tilt'' (t-) bands \cite{SM98,Wa93}, 
unlike the backbending caused by the two-bands crossing
of the g- and s-bands \cite{SS72} in light rare-earth nuclei.
The t-band was recently proposed as ``tilted rotational band'' 
\cite{Fr93}.
In nuclei in the $A\simeq 180$ region,
the Fermi level lies in states of high-$\Omega$
($\Omega$ is  a projection of single-particle angular momentum onto the
symmetry axis of a nucleus).
As a result, low-$\Omega$ states associated with the rotation-alignment
are mostly occupied. Therefore, it is natural to
 expect a new type of excited bands involving the high-$\Omega$ states.
Hereafter we refer to the bands as high-$K$ bands.
Such bands as $K^{\pi}=8^{+}$ ($10^{+}$) are actually
observed in the vicinity of the yrast line 
\footnote{ ``yrast line''  means a sequence of 
the lowest states in energy for given  angular momentum.}
in $^{180}$W and $^{182}$Os 
\cite{WY93,KL95} ($^{182}$W and $^{184}$Os \cite{SM95,SM98}).
It is characteristic that these bands have inter-band E2 transitions towards 
the yrast band having low-$K$ configurations, namely 
these transitions violate the $K$-selection rule to a great extent 
($\Delta K \simeq K$).
Walker et al. considered that these transitions are related to a 
new type of the backbending mechanism involving 
tilting degrees of freedom in the rotation axis \cite{WY93}.
The nature of the high-$K$ bands changes at lower and higher spin 
than that in the backbending region ($I_{\rm c}\simeq 14\hbar$).
At $I < I_{\rm c}$, the bands form a regular \delone sequence, 
while at $I > I_{\rm c}$ the bands split into two \deltwo sequences.
%
After the splitting, 
the odd-spin sequence becomes energetically lower than the even-spin one,
although the high-$K$ bands begin with even spin ($I=8\hbar$ or $10\hbar$).
This type of splitting is conventionally called ``signature inversion''.
These features in the high-$K$ bands suggest a certain strength of
inter-band interaction between the low- and high-$K$ bands. 

There are, at least, two modes that couple states having
intrinsic structures different in the $K$-quantum number.
One is the wobbling motion
that we present in this letter and the other is the triaxial deformation 
(i.e., sizeable $\gamma$ deformation).
%
%
In our recent calculations based on the two-dimensional cranked HFB (2d-CHFB)
\cite{AOO99}, 
the intrinsic states of $^{182, 184}$Os have (almost stable) prolate shape
($\gamma < 3^{\circ}; \beta \simeq 0.275$)
in the backbending region ($I \simeq I_{\rm c}$).
Also, the calculations show that tilting and $\gamma$
degrees of freedom are almost decoupled
in the  region, while
the coupling is
seen in the {\it very} high-spin region like $I\simeq 36\hbar$.
In this letter, we focus on the wobbling motion 
decoupled from the $\gamma$-degree of freedom in $I\simeq I_c$.

The conventional cranking model, which is restricted to
the rotation about the principal axis of the mass quadrupole deformation (PAR),
can be extended to the rotation about an axis tilted away
from the principal axis (TAR).
%
%
PAR simulates low-$K$ bands such as the g- or s-bands, 
while TAR describes high-$K$ bands .
In the present paper, both the PAR and TAR are calculated self-consistently 
through the 2d-CHFB method.
We regard the wobbling motion  as a dynamical mode to couple 
the low- and high-$K$ states in a quantal and microscopic approach.
We treat the motion in terms of
the generator coordinate method after the angular momentum projection ({\it 
GCM after AMP}),
in which tilt angle of the rotating axis is chosen to be
the generator coordinate.

In our previous paper \cite{HOO95}, 
we presented calculations  based on the wobbling model (GCM based on 2d-CHFB
wave functions)
{\it without} angular momentum projection and encountered a difficulty
in solving the Hill-Wheeler equation: a problem concerned with convergence
in the eigenvalues with respect to the ``cut-off'' dimension, which we
will discuss below.
We tried to improve the method 
through the constrained Hill-Wheeler equation,
but could not obtain satisfactory results \cite{HOOA}.
All the difficulties seem to come from the broken symmetries by the
mean field approximation. (Angular momentum and numbers of
nucleons are not conserved in  deformed HFB states.)
In many cases, the number constraints may be good enough to produce the
reasonable cranked HFB states of stable nuclei \cite{WAM84,RS80}.
We consider that, in a study of high-spin states,
restoration of the rotational symmetry is more
important than the gauge invariance,
as the first attempt to circumvent the difficulties.

The 2d-CHFB states are wave packets of eigenstates in angular momentum 
(and number)
and have broad width around the constrained value $J$ \cite{OOT98,RS80}.
Thus, the GCM states also include  substantial components of
undesirable angular momenta far from  $I=J$.
%
%
In order to eliminate them, we perform exact angular momentum projection (AMP) 
on 2d-CHFB states.
%
Norm and energy overlap
kernels are obtained through the formula \cite{OY66}.
But the well-known ambiguity of the phase by $\pi$ 
inheres in the formula for the norm overlap kernels.
We employ the analytic continuation method to determine the 
proper branches.
\cite{HHR82}.
%
%

The parameter set of the pairing+Q$\cdot$Q model
Hamiltonian employed here and details of the self-consistent 2d-CHFB method
are found in Ref.\cite{AOO99}. 
In the 3d-cranking model, constraints on three components of
angular momentum $J_{k}=\langle \Phi \mid \hat{J}_{k} \mid \Phi \rangle $
are specified by the tilting angle $(\theta,\phi)$ and
$J$ through
$J_{1}=J\cos\theta\cos\phi$, $J_{2}=J\cos\theta\sin\phi$
and $J_{3}=J\sin\theta$.
Here, a cranked HFB state is solved for $^{182}$Os and 
denoted as $|\Phi(\theta\phi;J)\rangle$.
The latitude ($\theta$) and longitude ($\phi$) angles are introduced
so as to measure the deviation of the rotating axis from the $x$-axis (PAR).

Now, we set up a wobbling state, or GCM state ($|\Psi\rangle$), which 
is described as the superposition of projected 
3d-cranked HFB states with different tilting angle:
\begin{equation}
|\Psi^I_M\rangle =\sum_{K=-I}^{I}\int^{\theta_0}_{-\theta_0} \! d\theta 
\int^{\phi_0}_{-\phi_0} \!d\phi \ 
f^I_K(\theta \phi)\hat{P}^I_{MK}
|\Phi(\theta \phi;J)\rangle,
\end{equation}
where $\hat{P}^I_{MK}$ is the angular momentum projection operator \cite{RS80}.
%
%
%
%
%
The function $f^I_K(\theta,\phi)$ is called the generator wave function.
The azimuthal angle $\phi_0$ is 
taken to be zero in the present calculations and is omitted hereafter.
As for $\theta_0$ on the integration limit, we will explain later.
$J$ is fixed to be $14\hbar$, and we also see later some properties of
the CHFB states with a constraint $J=14\hbar$ briefly.
The GCM state is obtained by the variational principle with 
respect to $f^I_K(\theta)$,
leading to the so-called Hill-Wheeler equation:
\begin{equation}
  \label{HWeq}
  \sum_{K'}\int d\theta' \left( H^I_{KK'}(\theta,\theta') -
E^IN^I_{KK'}(\theta,\theta') \right)f^I_{K'}(\theta')=0.
\end{equation}
The norm and energy overlap matrix-kernel $N^I_{KK'}(\theta,\theta')$ and 
$H^I_{KK'}(\theta,\theta')$ are respectively defined as follows,
\begin{equation}
  \label{OvKr1}
  \left(
    \begin{array}{c}
      N^I_{KK'}(\theta,\theta')\\
      H^I_{KK'}(\theta,\theta')
    \end{array}
  \right)
  =
  \left(
    \begin{array}{c}
      \langle\Phi(\theta)|\hat{P}^I_{KK'}|\Phi(\theta')\rangle \\
      \langle\Phi(\theta)|\hat{H}\hat{P}^I_{KK'}|\Phi(\theta')\rangle
    \end{array}
  \right).
\end{equation}
%
%
We solve Eq.(\ref{HWeq}) in two steps \cite{HOO95}: 
(1) Diagonalize the norm overlap 
matrix kernel like,
\begin{equation}
\label{Nmatrx}
\sum_{K'}\int_{-\theta_0}^{\theta_0} d\theta' N^I_{KK'}(\theta,\theta') \chi^{IK'}_{n}(\theta')
=\nu^I_{n}\chi^{IK}_{n}(\theta),
\end{equation}
to obtain eigenvalues 
$\nu^I_n$ and eigenfunctions $\chi^{IK}_n(\theta)$,
in which $n$ is an index specifying $\nu^I_{n}$ in order of magnitude.
The eigenvalues are never negative owing to properties of the norm overlap
matrix-kernels;
(2) Define  energy matrices on the base of orthonormal set 
$\mid nIM \rangle = \sum_{K}\int_{-\theta_0}^{\theta_0} d\theta 
\chi^{IK}_{n}(\theta)/\sqrt{\nu_n^I}
\hat{P}^I_{MK}\mid \Psi(\theta) \rangle $
\begin{equation}
  \label{Ematrix}
  {\cal H}^I_{nn'}=\sum_{KK'}\int\!\! \int d\theta d\theta'
\frac{\chi^{IK*}_n(\theta)}{\sqrt{\nu^I_n}}
H^I_{KK'}(\theta,\theta')
\frac{\chi^{IK'}_{n'}(\theta')}{\sqrt{\nu^I_{n'}}}.
\end{equation}
Eventually, the Hill-Wheeler equation gets transformed into an
eigenvalue equation: $\sum_{n'}{\cal H}_{nn'}g_{n'} = E^I g_{n}$. 
Integrations in Eq.(\ref{Ematrix})
are achieved  numerically by an approximation of discretization.
%
In the present study, five angular points are chosen 
($|\Phi(J,\theta)\rangle;
  \theta=0^{\circ},\pm 7^{\circ}$ and $\pm 20^{\circ}$)
under a condition:
$\langle\Phi(\theta_i)|\Phi(\theta_{i+1})\rangle < 0.9$, up to 
$|\theta_{\rm max}| =  20^{\circ}$.  
%
%
%
There are two reasons why we take only angles up to 20$^{\circ}$.
The first reason comes from a physical argument: 
according to the energy curve for the CHFB states with  $J=14\hbar$
(See the curve for $J=15\hbar$ in Fig.4(a) in Ref.\cite{AOO99}),
the first valley around $\theta=0^{\circ}$ has an edge at $\theta=20^{\circ}$.
Realization of the wobbling motion around PAR is thus achieved approximately 
by the superposition of the CHFB states corresponding to this first valley.
Although we should attempt to examine 
a global wobbling motion from $\theta=-90^{\circ}$ to $\theta=+90^{\circ}$,
the present framework is satisfactory to study the mechanism of mixing between
high- and low-$K$ states.
The second reason comes from numerical difficulty to calculate the norm
kernels at larger tilt angles which include quite high frequency modes. 
For a precise treatment of overlap kernels
we need much finer meshes in the Euler angles,
but this requirement in numerical calculations cannot be feasible
within an available computation time. 
%
%
%
%

The net dimension of the Hill-Wheeler equation  turns out to be 
$n_{\rm max} = \left\{ 5\times (2I+1)\right\}$ 
for a given angular momentum ($I$).
%
%
%
Fig.1 shows the spectrum of each band 
(the yrast, yrare, and second yrare band
\footnote{ ``yrare band''  means a sequence of 
the first excited states from the yrast states.}
)
as a function of $n_{\rm cut}$.
A  ``plateau''
indicates that the solutions (i.e., energy eigenvalues) are almost
independent of the cut-off dimension, $n_{\rm cut}$.
This result implies a great advantage of AMP 
in comparison with our previous GCM calculations.
%
%
%
%
%
From the results, we truncate at $n_{\rm cut}=11$ for all states
except the $I=17$ yrare state, in which we truncate at $n=10$.
The odd-$I$ states in the first yrare band form ``slopes'', 
{\it except} narrow plateaus formed around $n_{\rm cut}=10$.
The reason for the appearance of the ``slopes'' may 
come from mixing of the states
with the different particle number, i.e., ``contaminations in number'', 
for we have not carried out number projection in this work.
Though the ``contaminations'' make our discussion dubious,
the dual projection in number and angular momentum is
nothing but a heavy calculation consuming a great deal of computer time.
At present, we leave an examination on the effect of number projection 
as a future study.

Fig.2 shows the resultant energy spectrum in the band crossing region.
The signature splitting and inversion take place in the $I \ge 15\hbar$ region.
The phase of the staggering is properly reproduced comparing with the
experiment: odd-$I$ states are energetically lower than even-$I$ states.
However, the amount of the splitting is much larger than experimental values.
Our result is roughly 500 keV while the experiment is about 100 keV.
There are several reasons for this discrepancy. 
There may be some argument for a choice of 
the pairing-plus-quadrupole force, but a more important factor is a set of 
parameters for the effective force. The parameters should be adjusted 
for calculations of the {\it GCM after AMP} states, though this procedure
is again difficult because of the limit of the present computation resources.
Thus, we have to keep using the same parameters as in the HFB calculations.
Other possibility for the discrepancy is that we take a fixed $J$ value 
($J=14\hbar$) in the CHFB states. 
For example, in this study, 
the projected states with $I=17\hbar$ are projected out
from the CHFB state with $J=14\hbar$. 
To make a better projected states,
it can be better to project out from the CHFB states with $J=17\hbar$.

It is satisfactory  that the \delone band sandwiched between the
two \deltwo bands is reproduced\footnote{We pick up the lowest three 
levels out of 11 levels in order to draw the Fig.2.} in Fig.2. 
We are tempted to say this band structure is produced via multi-bands
crossing among g, s, and high-$K$ bands.
However, we should note that 
the higher \deltwo band does not have a g-band character but a s-band one.
This result can be explained as below:
we have fixed the angular momentum constraint on $J=14\hbar$, and
the gap energy for neutron in this range of angular momentum 
is substantially reduced  due to the rotation-alignment 
(Refer to the result in Ref.\cite{AOO99}).  
Therefore, our generating states have a good amount of the rotation-aligned 
component, i.e. a strong character of the s-band.
Consequently, the upper \deltwo band is of the s-band character
instead of the g-band.

Backbending is not well reproduced in the present calculations.
We can suppose that this deficiency is also due to the lack of the g-band
components  in our calculations because
the g-band can contribute to up- or back- bending  through its crossing with
the s-band.
To make a complete analysis of the three-bands crossing in $^{182}$Os,
we need to use, at least, two different CHFB states: one with low $J$ and
the other with high $J$, and perform the {\it GCM after AMP} 
calculations around at each angular momentum. 
However, for a  practical reason, 
the present paper is dedicated to an intensive study 
of the mixing of low- and high-$K$ 
states through the wobbling motion, which corresponds to the 
band crossing of the s-band with the t-band. 
This situation can happen just after the g-s crossing.
As for the g-t crossing, we will discuss later.

Let us learn the obtained $K$ distributions of the wave functions
in the context of the interaction between the lowest two bands.
For this purpose, we calculate overlaps between the wobbling (GCM) state and
each spin-projected intrinsic state ( i.e., 2d-CHFB states):
$\langle\Phi(J,\theta)|\hat{P}^{I\dag}_{MK}|\Psi^I_{M}\rangle 
\equiv g^I_{K}(\theta)$ \cite{RS80},
%
%
 evaluated by using the formula,
\begin{equation}
g^I_K(\theta)=\sum_{n=1}^{n_{\rm cut}}\sqrt{\nu_n}g_n\chi_{n}^{IK}(\theta).
\end{equation}
Fig.3 shows the quantity $|g^I_K(\theta)|^2$ with respect to  $K$ 
for each $\theta$ value.
In our calculations, the $I=14\hbar$ and $I=16\hbar$ states 
correspond to the states before and after the crossing, respectively.
In $I=14\hbar$, 
the yrast state shows a typical feature of the s-band:
the major component is $K=0$ and some fluctuations  around it, 
i.e., $\frac{|\Delta K|}{2}\simeq 1\hbar$ or $2\hbar $.
This feature implies that 
a good amount of the rotation-aligned components are induced
in the 2d-CHFB states at $J=14\hbar$,
.
In turn, the yrare state
has two large peaks at $K=\pm 8\hbar$, and one
small peak at $K=0\hbar$. These features indicate a typical
high-$K$ band. Most of the high-$K$ components are brought about  by
TAR states ($ |\theta |=20^{\circ} $).
The existence of the $K=0$ component implies the 
inter-band interaction with the yrast band, though the interaction is
weak due to the large difference in energy
\cite{AOO99}.
In the crossing region, one can expect that 
the mixing is supposed to be stronger
as the two bands come closer. 
In $I=16\hbar$,  we can find that both of the yrast and yrare states
 have three peaks at $\pm 8\hbar$ and $K=0$,
indicating that the bands are mixed with each other
after the crossing.
In particular, the yrare states have the three comparable peaks 
at low- and high-$K$ components.
In these perturbed states,
the high-$K$ components come mainly from TAR states while the 
low-$K$ components are brought by PAR states.
Hence, these states are interpreted as the wobbling states,
namely, dynamically rotating states coupling the low-$K$ PAR
and high-$K$ TAR states.
On the other hand, odd-$I$ states in the high-$K$ band
are much {\it less perturbed} according to the calculations.
This is 
because the odd-$I$ members are energetically isolated
and have no partners lying nearby to couple  ( PAR has only even-$I$ states).

This fact draws an interpretation for the signature inversion 
in the t-bands:
The inter-band interaction between the s- and t-bands pushes up (down)
the even-$I$ states in the t-band (s-band). 
(Note that the s-band is the yrast after the crossing with the g-band.)
This downward shift of the yrast band may enhance backbending,
though it is not clearly seen in the present calculation.
The interaction between the g- and t-bands
is quite weak because the g-band is approximately the pure $K=0$ state 
with small admixture of high-$K$ components.
As a consequence,
the backbending in the $A\simeq 180$ region 
occurs not only due to the rotation-alignment mechanism dominating in the rare-earth, but also due to the mechanism of the wobbling motion, i.e., 
the low- and high-$K$ interaction.
On the other hand, the odd-$I$ states are unperturbed and 
serve as reference states for the interaction energy. 

In summary,
we have investigated a mechanism of backbending and the signature inversion
in $^{182}$Os,
by means of {\it GCM after AMP}, on the  2d-cranked HFB states.
With this method, 
we have qualitatively reproduced the main feature of 
the level structure showing
the signature inversion in the high-$K$ band lying very closely to the
yrast band. 
We interpret this result from a point of view of 
an inter-band interaction between the low-$K$ and high-$K$ bands.
We have shown that the perturbed states have 
characters of the wobbling motion, that is, dynamical mode 
coupling low-$K$ PAR and high-$K$ TAR states.
In terms of this wobbling model, we have discussed
an enhancement of the backbending in the $A\simeq 180$ region, 
in which the typical rotation-alignment is somewhat suppressed 
due to the location of the Fermi level.
It is also shown that AMP brings about a great advantage in stabilization
of eigenvalues in the Hill-Wheeler equation.
At the same time, we realize that number projection turns out be
important since it may eliminate a ``slope'' owing to
the ``number contamination'', 
as we expand the GCM states with the orthonormal basis. 

%

We would like to thank Dr. N. Tajima for discussions and suggestions
on this work.
%
%
Most of the computations have been done using the 
Vector Parallel Processor, Fujitsu VPP500/28 at RIKEN which 
are also gratefully acknowledged.

\newpage
\begin{center}
\Large \bf Figure Captions
\end{center}


\noindent
{\bf Fig.1 :}
A relation between the cut-off dimensions ($n_{\rm cut}$)
and the energy eigenvalues ($E^I(n_{\rm cut})$)
, in the Hill-Wheeler equation.
The ``plateaus'' of the graphs imply the possibility of ``safety cut-off''
within the plateaus. In the present study, we choose $n_{\rm cut}=10$ for
the first yrare $I=17\hbar$ state and $n_{\rm cut}=11$ for the other
states.\\

\noindent
{\bf Fig.2 :}
Energy spectrum calculated through the {\it GCM after AMP}.
The angular momentum constraint for the intrinsic state is fixed to be
$J=14\hbar$.
From the analysis of the generator wave functions,
the yrast and the second yrare bands (\deltwo-bands) 
are of the low-$K$ characters,
while the first yrare band (\delone-band) is of the high-$K$ character.
The signature inversion is seen at the region $I > 15\hbar$.
\\

\noindent
{\bf Fig.3 :}
Graphs for the overlaps between the GCM state and the 
spin-projected intrinsic state:
$g^I_K(\theta) 
\equiv \langle \Psi(J,\theta)|\hat{P}^{I\dag}_{MK}|\Psi^I_M\rangle$.
(See texts.)
The left panels are for the GCM states before the 
band crossing ($I=14\hbar$), and
the right panels are for the states after the crossing ($I=16\hbar$).
The lower and upper panels show the yrast and first yrare states,
respectively. In each panel, the graphs are separately drawn for 
each tilt angle: $\theta=0^{\circ},\pm 7^{\circ},$ and $\pm 20^{\circ}$.
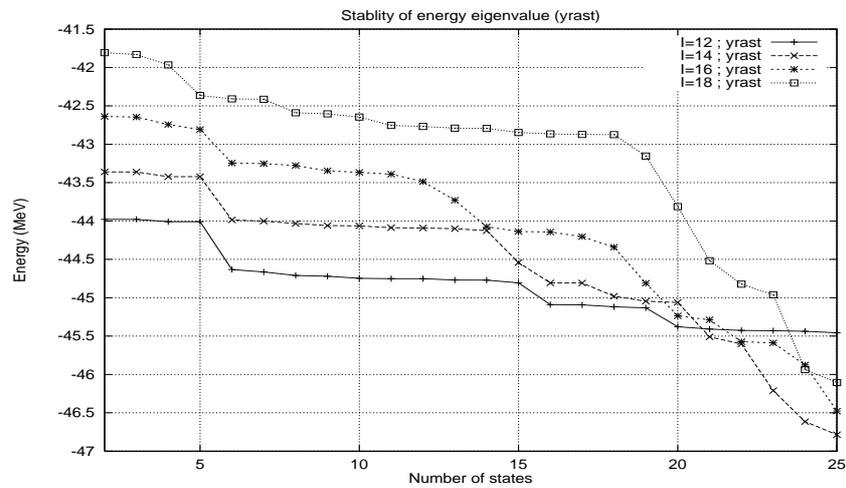
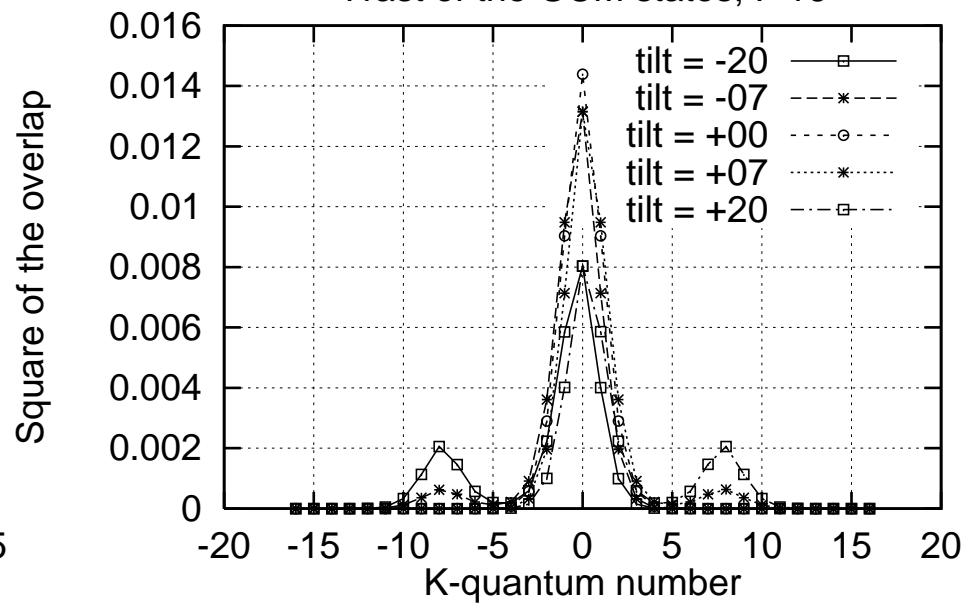
\end{document}